\documentclass[preprintnumbers,amsmath,amssymb,superscriptaddress]{revtex4}
\usepackage{epsfig}
\usepackage{psfrag}
\usepackage{amsfonts}
\usepackage{graphicx}% Include figure files
\usepackage{dcolumn}% Align table columns on decimal point
\usepackage{bm}% bold math
\usepackage{mathtools,amssymb,bm}
\usepackage{graphicx}
\usepackage[caption=false]{subfig}
\begin{document}

%\preprint{APS/123-QED}

\title{Probing the inhomogeneous phases in magnetized quark matter via the bosonic two-point function }% Force line breaks with \\

\author{Kang Deng}

\author{Bo  Feng}

\author{Tianle Wei}

\author{Shuai Yuan}

\affiliation{
School of Physics, Huazhong University of Science and Technology, Wuhan 430074, China
}

%\author{Efrain J. Ferrer}
%\affiliation{Dept. of Physics and Astronomy, Univ. of Texas Rio Grande Valley, 1201 West University Dr., Edinburg, TX 78539, USA}

%\author{Hai-cang Ren}
%\affiliation{
%Physics department, The Rockefeller University, 1230 York Avenue, New York, New York 10021-6399, USA}
%\affiliation{
%Institute of Particle Physics and Key Laboratory of Quark and Lepton Physics
%(MOE),  Central China Normal University, Wuhan 430079, China}

%\date{\today}% It is always \today, today,
             %  but any date may be explicitly specified

\begin{abstract}

We investigate the stability of homogeneous chiral condensates in magnetized quark matter within the two-flavor Nambu-Jona-Lasinio model. By expanding the effective action to quadratic order, we derive the full momentum-dependent bosonic two-point function in a constant magnetic field. At zero temperature, our analytical results show that each occupied Landau level produces a sharp logarithmic divergence at its corresponding Fermi momentum. This severe Peierls instability is driven by one-dimensional Fermi surface nesting within the effectively 1+1-dimensional subsystems created by Landau quantization. Numerically, we map out the momentum-space topography of the two-point function across varying magnetic fields and temperatures. We find that strong magnetic fields induce deep negative canyons. While thermal fluctuations smooth these canyons into shallower valleys as the temperature increases, the negative regions remain robust at intermediate temperatures and only vanish entirely at sufficiently high temperatures.  Intriguingly, at larger chemical potentials where the system is strictly stable, a broad positive basin with a flat curvature at the origin persists at non-zero momentum. This non-monotonic dispersion is the hallmark of a moat regime, acting as a precursor state that strongly signals a residual tendency toward inhomogeneous chiral symmetry breaking. Our findings provide valuable insights into the phase structure of dense, magnetized quark matter.
\end{abstract}

%\pacs{74.20.Fg,03.75.Nt,11.10.Wx,12.38.-t}% PACS, the Physics and Astronomy
                             % Classification Scheme.
%\keywords{Suggested keywords}%Use showkeys class option if keyword
                              %display desired
\maketitle

% 74.20.Fg BCS theory and its development
%03.75.Nt Other Bose-Einstein condensation phenomena
%11.10.Wx Finite-temperature field theory
%12.38.-t Quantum chromodynamics (for quarks, gluons, and QCD in nuclear reactions, see 24.85.+p)

%%%%%%%%%%%%%%%%%%%%%%%%%%%%%%%%%%%%%
\section{Introduction}
%%%%%%%%%%%%%%%%%%%%%%%%%%%%%%%%%%%%%

Understanding the phase structure of Quantum Chromodynamics (QCD) at finite temperature and baryon chemical potential remains one of the most challenging and fascinating problems in modern high energy nuclear physics \cite{FukushimaHatsuda2011}. While the high-temperature regime is accessible through lattice Monte Carlo simulations \cite{Ding_lattice} and heavy-ion collision experiments, the region of low temperature and high chemical potential, relevant for the interior of neutron stars and neutron star mergers, suffers from the severe fermion sign problem that renders lattice methods inapplicable. In this regime, complementary approaches, including effective field theories \cite{Du}, functional methods \cite{Fischer_effective_qcd}, and model studies \cite{Buballa_NJL,Klevansky-Rep}, are indispensable for providing theoretical insights.

Among the most intriguing possibilities for the high-density region in the QCD phase diagram is the emergence of inhomogeneous or "crystalline" phases, in which the chiral condensate, the order parameter for chiral symmetry breaking, exhibits periodic spatial modulation. Such phases simultaneously break chiral symmetry and translational invariance, leading to a rich ground state structure reminiscent of spin-density waves in condensed matter systems \cite{Gruner, FF}. The concept of density waves in strongly interacting matter has a long history, originating from early work on nuclear matter \cite{Overhauser} and subsequent investigations of pion condensation \cite{Migal}. Shortly thereafter, the first relativistic treatment of chiral density waves was developed \cite{Relativistic_density_wave}, and the idea was eventually extended to quark matter \cite{Inhomo_QM_1, Inhomo_QM_2}. A major breakthrough came with the exact analytical solution of the 1+1 dimensional Gross-Neveu model in the large-N limit \cite{Thies}, which rigorously established the existence of an inhomogeneous phase and revealed the exact form of the modulating condensate in terms of Jacobi elliptic functions. These exact results demonstrated that inhomogeneous phases are not merely theoretical curiosities but can be rigorously realized in relativistic quantum field theories. Inspired by these findings, extensive investigations were carried out in higher-dimensional models, particularly the Nambu-Jona-Lasinio (NJL) model \cite{Nickel, DCDW, Buballa} and the Quark-Meson (QM) model \cite{Tripolt,Carignano2014,Adhikari2017}, which serve as low-energy effective theories of QCD. In these models, inhomogeneous phases typically appear at low temperatures and intermediate chemical potentials, covering the region where homogeneous model studies predict a first-order phase transition.

A fundamental difficulty in investigating inhomogeneous phases lies in the fact that the spatial modulation pattern of the condensate is not known a priori, leaving in principle infinitely many possibilities. Two complementary approaches have been developed to address this challenge. The first involves making an explicit ansatz for the spatial modulation such as the chiral density wave or the real-kink crystal, and comparing the free energy of this configuration with that of homogeneous phases. While conceptually straightforward and widely used, this approach risks missing energetically preferred solutions that are not captured by the chosen ansatz. The second approach, the stability analysis, offers a complementary perspective by examining whether a spatially homogeneous ground state is unstable against small inhomogeneous perturbations of arbitrary shape \cite{Koenigstein2022,Pannullo2022,Pannullo2023a,Pannullo2023b,Buballa2019,Carignano2020}. By expanding the effective action around the homogeneous configuration and analyzing the curvature in field space, this method provides a shape-agnostic probe of the system's tendency toward spatial modulation. A negative curvature at some finite momentum signals an instability, indicating that the true ground state must be inhomogeneous. The key advantage of this approach is that it does not require prior knowledge of the modulating pattern, making it a powerful and versatile tool for exploring the QCD phase diagram.

However, the model results are not without controversy. Recent studies have shown that in 2+1 dimensions, inhomogeneous phases vanish when the continuum limit is taken \cite{Pannullo2022,Buballa2021}, and that their existence in the 3+1 dimensional NJL model depends sensitively on the regularization scheme \cite{Pannullo2023b,Pannullo2024}. These findings raise important questions about whether inhomogeneous phases are genuine physical phenomena or merely artifacts of model truncations and regulator choices. To resolve these questions, it is essential to develop methods that can be applied directly to QCD, without relying on the approximations inherent in effective models.

Extending the stability analysis to full QCD, however, presents formidable challenges. Unlike in mean-field models, the quark self-energy in QCD is non-local due to gluon-mediated interactions, and the free energy cannot be written as a simple functional of local condensates. In a series of recent works \cite{Motta2023,Motta2024,Motta2024b,Motta2025}, the authors have developed a generalization of the stability analysis based on the two-particle irreducible (2PI) effective action that is applicable to any truncation of QCD. This framework introduces test functions for inhomogeneous perturbations of the quark propagator and extracts the stability condition by analyzing the second-order variation of the effective action. The method has been successfully tested in a rainbow-ladder truncation of QCD with various models for the effective running coupling, revealing clear evidence that the chirally symmetric solution becomes unstable against inhomogeneous perturbations below a certain temperature \cite{Motta2024}. However, this approach relies on the explicit choice of test functions for the perturbations, which can be computationally expensive and constraint-dependent. To overcome this, the same authors further proposed the inhomogeneous chiral susceptibility as a novel and test-function-free tool to detect spatially modulated chiral symmetry breaking \cite{Motta2024b}.

%Most recently, a novel and computationally efficient approach has been proposed %based on the concept of an inhomogeneous chiral susceptibility [27]. By applying a %chiral rotation to the quark fields, the problem of inhomogeneous condensation is %mapped onto a homogeneous problem in the rotated basis, at the cost of %introducing an effective momentum-dependent term in the propagator. The response %of the system to this term determines whether inhomogeneous chiral symmetry %breaking occurs. This method not only reproduces the known NJL model results but %also confirms and extends the earlier QCD stability analysis, definitively delineating %the region where the homogeneous chiral solution is unstable toward crystallization.

Despite these advances, several important questions remain open. Quantum fluctuations, particularly those of Nambu-Goldstone bosons associated with broken symmetries, may disorder long-range crystalline order, potentially leading to liquid-crystal phases or quantum pion liquids rather than true condensates \cite{Pisarski2019, Pisarski2020}. Even if fully ordered phases are disfavored, the underlying tendency toward spatial modulation may manifest as "moat regimes" with non-monotonic meson dispersion relations, exhibiting minima at non-zero momentum. Such features could leave detectable signatures in heavy-ion collision experiments \cite{Pisarski2021,Rennecke2023,Fukushima2024,Fu2025,Nussinov2024}. Importantly, moat regimes can be understood as precursors of inhomogeneous phases. The characteristic wavevector at which the dispersion minimum occurs corresponds precisely to the modulation scale around which an inhomogeneous condensate would form. The stability analysis provides a direct probe of such tendencies by identifying the momentum scale at which the homogeneous ground state becomes unstable.

Beyond the challenges posed by finite baryon chemical potential, another crucial dimension of the QCD phase diagram is the presence of strong external magnetic fields \cite{Andersen2016, Endrodi2025}. Such fields are realized in nature in the most extreme astrophysical environments. Magnetars, neutron stars with surface magnetic fields reaching $10^{14}\sim 10^{15}$G and interior fields possibly orders of magnitude higher, provide a unique laboratory for studying the interplay between magnetic fields and strong interactions. Moreover, in non-central heavy-ion collisions at facilities such as RHIC and the LHC, transient magnetic fields of comparable strength are generated, offering the possibility to probe magnetized QCD matter in terrestrial experiments.

%The influence of magnetic fields on the QCD vacuum has been a subject of intensive study for decades. A %key phenomenon is magnetic catalysis—the enhancement of chiral symmetry breaking by an external %magnetic field, first observed in the context of the Gross-Neveu model [35,36] and subsequently established %in QCD-inspired models and low-energy effective theories [37,38]. In essence, a magnetic field reduces the %dimensionality of fermion dynamics in the infrared, strengthening the attractive interaction that binds %quark–antiquark pairs and thereby increasing the chiral condensate. This effect has profound implications for %the phase diagram: the critical temperature for chiral restoration is shifted upward, and the nature of the %phase transition may be altered.

The presence of a magnetic field breaks rotational symmetry and introduces additional scales, enriching the possible modulation patterns. For instance, the chiral density wave may acquire a preferred direction aligned with the magnetic field, leading to the so-called magnetic dual chiral density wave (MDCDW) phase (see the review \cite{Efrain_review} and references therein), which has been explored in the context of the NJL model \cite{Frolov2010,Ferrer2018,Ferrer2020,Feng2020}. These investigations suggest that the magnetic field can not only support inhomogeneous order but also protect it against certain types of fluctuations that would otherwise destroy long-range crystalline order \cite{Ferrer2020}. This protective effect is particularly intriguing because it offers a potential resolution to the Landau-Peierls instability, which plagues one-dimensional modulations in three spatial dimensions. Despite these promising indications, a systematic and model-independent analysis of inhomogeneous phases in magnetized QCD is still lacking. The stability analysis framework developed in \cite{Koenigstein2022,Pannullo2022,Pannullo2023a,Pannullo2023b,Buballa2019,Carignano2020} can be adapted to this setting. Such an endeavor is essential to establish robust predictions for the behavior of strongly interacting matter under the extreme conditions encountered in neutron stars and heavy-ion collisions, and to guide experimental searches for signatures of crystalline order in the QCD phase diagram. To fill this gap, in this paper we perform a stability analysis of the homogeneous chiral condensate in the two-flavor NJL model by deriving the full momentum-dependent bosonic two-point function in a constant magnetic field and mapping out its momentum-space topography.

The remainder of this paper is organized as follows. In Section~\ref{sec:NJL_model}, we establish the theoretical framework by introducing the magnetized NJL model and determining its homogeneous chiral condensate. Section~\ref{sec:stability} is devoted to deriving the scalar bosonic two-point function, which serves as our stability criterion for spatially modulated perturbations. Building on this formalism, Section~\ref{sec:numerical_analysis} investigates the momentum-dependent structures of the two-point function. Specifically, Section~\ref{subsec:peierls} provides an analytical derivation of the zero-temperature Peierls instability driven by one-dimensional Fermi surface nesting, while Section~\ref{subsec:num_results} presents the numerical momentum-space topography, including the finite-temperature evolution and the emergence of a moat regime. Finally, we summarize our findings in Section~\ref{sec:conclusion}, with detailed derivations of the two-point function collected in Appendix~\ref{app:Gamma2_derivation}.

%%%%%%%%%%%%%%%%%%%%%%%%%%%%%%%%%%%%%
\section{Mean-field Effective Potential and the Homogeneous Condensate}
%%%%%%%%%%%%%%%%%%%%%%%%%%%%%%%%%%%%%
\label{sec:NJL_model}

We consider a two-flavor Nambu--Jona-Lasinio (NJL) model of massless quarks in the presence of an external, constant, and uniform magnetic field. The Lagrangian reads
\begin{equation}
\mathcal{L} = \bar{\psi}\left[i\gamma^{\mu}(\partial_{\mu} + iQA_{\mu}) + \mu \gamma^{0}\right]\psi + G\left[(\bar{\psi}\psi)^{2} + (\bar{\psi} i\gamma^{5}\boldsymbol{\tau} \psi)^{2}\right],
\label{eq:Lagrangian}
\end{equation}
with $\psi^{T} = (u, d)$ the quark flavor doublet and $G$ the coupling constant of the four-fermion interaction. Here, $\boldsymbol{\tau}$ are the Pauli matrices and $Q = \mathrm{diag}(q_u, q_d)$ is the charge matrix in flavor space, with the flavor eigenvalues $q_u = 2e/3$ and $q_d = -e/3$. We take the magnetic field $B > 0$ along the $z$-axis and use the Landau gauge $A_{\mu} = (0, 0, Bx, 0)$. At $B = 0$, the interaction possesses the usual $\mathrm{SU}(2)_L \times \mathrm{SU}(2)_R$ chiral symmetry. The fixed electromagnetic background reduces the flavor symmetry because the two quark charges differ.

Introducing auxiliary fields $\sigma$ and $\boldsymbol{\pi}$ by a Hubbard-Stratonovich transformation, the partition function can be written as
\begin{equation}
Z =\int{\cal D}\psi{\cal D}{\bar\psi}\mathcal{D}\sigma \mathcal{D}\boldsymbol{\pi} e^{S},
\end{equation}
with
\begin{equation}
S[\bar{\psi}, \psi, \sigma, \boldsymbol{\pi}] = -\int_0^\beta d\tau\int d^3{\bf x} \left[\bar{\psi}\left(i\gamma^{\mu}(\partial_{\mu}+iQ A_{\mu}) + \mu \gamma_{0} - \sigma - i\gamma_{5}\boldsymbol{\tau}\cdot \boldsymbol{\pi}\right)\psi + \frac{\sigma^{2} + \boldsymbol{\pi}^{2}}{4G}\right].
\label{eq:action}
\end{equation}
The scalar auxiliary fields obeys $\langle \bar{\psi}\psi \rangle = -\langle \sigma \rangle /(2G)$ and $\langle \bar{\psi}i\gamma^{5}\boldsymbol{\tau}\psi \rangle = -\langle \boldsymbol{\pi}\rangle /(2G)$. Integrating out the quark fields yields the effective action
\begin{equation}
{\cal S}_{\rm eff}=-\int_{0}^{\beta}d\tau \int d^{3}{\bf x} \left( \frac{\sigma^{2} + \boldsymbol{\pi}^{2}}{4G} \right)+\mathrm{Tr}\ln D^{-1}[\sigma, \boldsymbol{\pi}], 
\end{equation}
with $D^{-1}[\sigma, \boldsymbol{\pi}] = i\gamma^{\mu}(\partial_{\mu} + iQ A_{\mu}) + \mu \gamma_{0} - \sigma - i\gamma_{5}\boldsymbol{\tau}\cdot \boldsymbol{\pi}$ and $\beta = 1/T$. The functional trace is taken over space-time, Dirac, color, and flavor indices.

In the subsequent analysis, we restrict the evaluated configurations and perturbations to the scalar channel by setting $\boldsymbol{\pi} = 0$. In this sector, the flavor-$f$ Dirac operator is $D^{-1}_{f} = i\gamma^{\mu}(\partial_{\mu} + iq_{f}A_{\mu}) + \mu \gamma_{0} - \sigma$, and the effective action becomes a functional of the scalar field $\sigma$.  To determine the mean-field equilibrium state, we assume a spatially homogeneous condensate denoted by $\sigma = \bar{\sigma}$. The corresponding homogeneous effective potential is defined by evaluating the effective action density at this constant background
\begin{equation}
U(\bar{\sigma},\mu,T,B) = -\frac{1}{\beta V} {\cal S}_{\rm eff}\Big|_{\sigma=\bar{\sigma}},
\label{eq:U_def}
\end{equation}
where $V$ denotes the spatial volume. The eigenvalues of the Dirac operator in the constant magnetic background are well known. We thus have
\begin{equation}
U(\bar{\sigma},\mu,T,B) = \frac{\bar{\sigma}^{2}}{4G} - N_c\sum_{f=u,d}\frac{|q_{f}|B}{\beta}\sum_{n=-\infty}^{\infty}\sum_{l=0}^{\infty}\alpha_l \int_{-\infty}^{\infty}\frac{dp_z}{(2\pi)^{2}} \ln \left[(\omega_n + i\mu)^{2} + E_{f,l}^{2}(p_z)\right],
\label{eq:U_LLL}
\end{equation}
where $\alpha_l = 2 - \delta_{l,0}$ accounts for the spin degeneracy of the Landau levels, $\omega_n = (2n+1)\pi T$ are the fermionic Matsubara frequencies, and $E_{f,l}(p_z) = \sqrt{p_z^2 + 2|q_f|Bl + \bar{\sigma}^2}$ is the energy of the $l$-th Landau level. 

The Matsubara sums can be carried out with standard techniques, yielding
\begin{equation}
U(\bar{\sigma},\mu,T,B) = \frac{\bar{\sigma}^{2}}{4G} - N_{c}\sum_{f=u,d}|q_{f}|B \sum_{l=0}^{\infty}\alpha_l \int_{-\infty}^{\infty}\frac{dp_z}{(2\pi)^{2}} \left[E_{f,l} + T\ln \left(1 + e^{-\beta (E_{f,l} + \mu)}\right) + T\ln \left(1 + e^{-\beta (E_{f,l} - \mu)}\right)\right].
\label{eq:U_thermal}
\end{equation}
The homogeneous condensate is determined by minimizing the effective potential, which yields the gap equation
\begin{equation}
\frac{\partial U}{\partial\bar{\sigma}} = 0.
\label{eq:gap_condition}
\end{equation}
Evaluating this derivative yields
\begin{equation}
\frac{1}{2G} - {N_c} \sum_{f=u,d}|q_f|B \sum_{l=0}^{\infty} \alpha_l  \int_{-\infty}^{\infty} \frac{dp_z}{(2\pi)^2} \frac{1}{E_{f,l}(p_z)} \left[1 - \eta(E_{f,l}(p_z)) - \bar{\eta}(E_{f,l}(p_z))\right] = 0,
\label{eq:gap_eq}
\end{equation}
where we have defined the Fermi-Dirac distributions for particles and anti-particles as
\begin{equation}
\eta(E) = \frac{1}{e^{\beta(E - \mu)} + 1}, \qquad \bar{\eta}(E) = \frac{1}{e^{\beta(E + \mu)} + 1}.
\label{eq:FD}
\end{equation}
The vacuum contribution in eq.~\eqref{eq:gap_eq} is ultraviolet divergent. We regularize it using a smooth momentum cutoff function
\begin{equation}
R_{\alpha}(p) = \frac{1}{2}\left[1 - \tanh \left(\frac{p - \Lambda}{\alpha}\right)\right],
\label{eq:regulator}
\end{equation}
where $p=\sqrt{2|q_f|Bl+p_z^2}$ represents the magnitude of the spatial momentum. For the numerical calculations, we retain the parameter values $\Lambda = 653$ MeV and $G = 5.02$ GeV$^{-2}$, originally fixed in the sharp three-momentum cutoff scheme by fitting the pion decay constant and the vacuum chiral condensate per flavor to $f_{\pi} = 93$ MeV and $\langle \bar{q} q\rangle = -(250 \mathrm{MeV})^{3}$ \cite{Buballa_NJL,Klevansky-Rep}, respectively. These parameters are retained without refitting when introducing the smooth regulator, with $\alpha = 0.01 \Lambda$. The same regularization prescription and parameter set are used consistently in the homogeneous gap equation and the bosonic two-point function.

\begin{figure*}[t]
	\centering
	\subfloat[\label{fig:homo_conden}]{%
		\begin{minipage}[b]{0.4\textwidth}
			\centering
			\includegraphics[width=\linewidth]{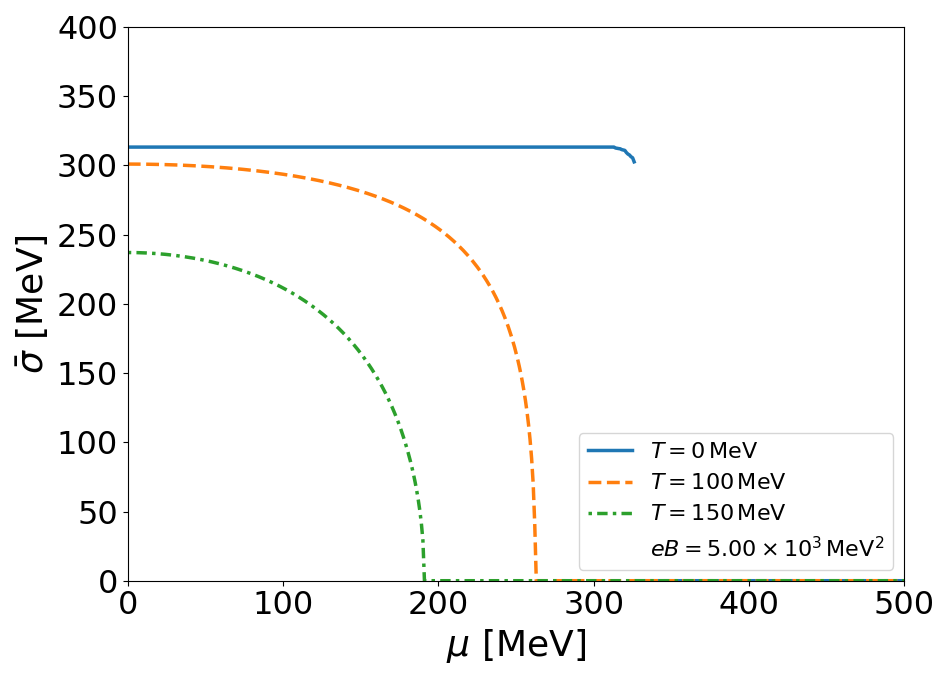}
		\end{minipage}%
	}
	\hspace{2cm}
	\subfloat[\label{fig:homo_conden_2}]{%
		\begin{minipage}[b]{0.4\textwidth}
			\centering
			\includegraphics[width=\linewidth]{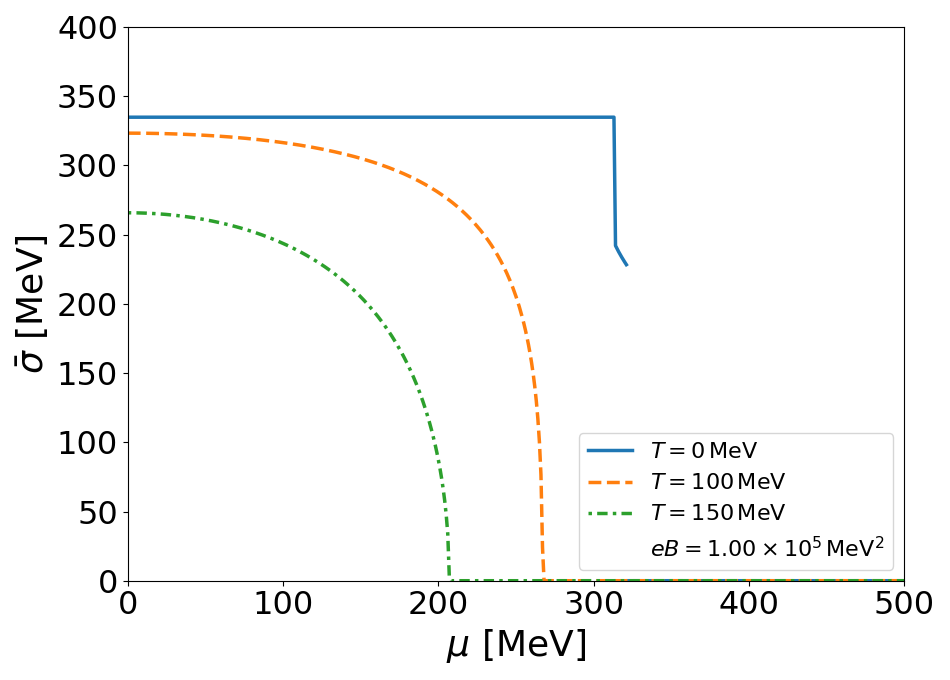}
		\end{minipage}%
	}
	\caption{Homogeneous condensate $\bar\sigma$ as a function of the quark chemical potential at $T=0$, $100$ MeV and $150$ MeV. The magnetic fields are (a) $eB=5\times10^3$ MeV$^2$, (b) $10^5$ MeV$^2$. }
	\label{fig:comparison}
\end{figure*}

Figs.~\ref{fig:homo_conden} and \ref{fig:homo_conden_2} display the homogeneous condensate obtained by minimizing the effective potential. At $T = 0$, the nonzero solution forms a plateau followed by a sharp drop to $\bar{\sigma} = 0$, indicating a first-order phase transition to the symmetric phase. In contrast, at nonzero temperatures, the condensate decreases over a broader chemical-potential range before reaching the symmetric phase. 

The zero-temperature plateau arises directly from the distribution functions in the gap equation, whose limits as $T \to 0$ are $\eta(E) = \Theta(\mu - E)$ and $\bar{\eta}(E) = 0$. Because $E_{f,l}(p_z) \geq |\bar{\sigma}|$, the medium contribution vanishes for $0 < \mu < \bar{\sigma}$, rendering the independence of the condensate on $\mu$. At finite temperature, these sharp step functions are replaced by smooth Fermi-Dirac distributions. Consequently, thermally occupied states contribute even when $\mu$ lies below the zero-temperature threshold, causing the condensate to decrease earlier than it does at $T = 0$. For sufficiently high temperatures, the phase transition becomes continuous. Furthermore, the condensate value at zero chemical potential increases at larger magnetic fields. This enhancement is consistent with the phenomenon known as magnetic catalysis of chiral symmetry breaking \cite{Klimenko1991,Gusynin1994,Shovkovy2013,BF_1,BF_2}.

%%%%%%%%%%%%%%%%%%%%%%%%%%%%%%%%%%%%%
\section{Bosonic two-point function}
%%%%%%%%%%%%%%%%%%%%%%%%%%%%%%%%%%%%%
\label{sec:stability}

%he homogeneous solutions provide the reference states for the subsequent stability %analysis. In particular, the vanishing homogeneous mass above the restoration transition %does not establish stability against spatial modulation. That question is addressed by the %momentum-dependent scalar two-point function: a negative value at nonzero momentum %identifies a spatially varying scalar deformation that lowers the free energy of the %homogeneous reference state.

The homogeneous solutions serve as the reference states for our subsequent stability analysis. Crucially, the vanishing of the homogeneous order parameter above the restoration transition does not guarantee stability against spatial modulations; lower-energy inhomogeneous phases may still exist. To investigate this possibility, we examine the momentum-dependent scalar two-point function. A negative value of this two-point function at nonzero momentum identifies an unstable mode, signaling a spatially varying scalar deformation that lowers the free energy relative to the homogeneous state. However, it should be emphasized that this stability analysis, which relies on a functional Taylor expansion, is only reliable in the symmetric restored phase where $\bar{\sigma} = 0$. As demonstrated in Ref.~\cite{Koenigstein2022}, in the broken phase ($\bar{\sigma}\ne 0$), the homogeneous minima are no longer saddle points with an unstable direction in momentum space when studying inhomogeneous perturbations. Consequently, they become true local minima. The genuine global minimum, i.e., the spatially oscillating phase, is separated from these homogeneous local minima by a substantial potential barrier. This barrier requires the order parameter to undergo large-amplitude excursions to traverse the high-energy region at $\bar{\sigma}=0$. Such large-amplitude oscillations violate the small-perturbation assumption of the Taylor expansion around a single homogeneous minimum, rendering the stability analysis ineffective in this regime.

To derive the two-point function, we expand the effective action about a homogeneous stationary background $\bar{\sigma}$, introducing a spatially varying perturbation $\delta\sigma(\mathbf{x})$ such that $\sigma(\mathbf{x}) = \bar{\sigma} + \delta\sigma(\mathbf{x})$. Denoting the Dirac operator evaluated at the homogeneous background as $\bar{D}^{-1} \equiv D^{-1}[\bar{\sigma}]$, we factor out $\bar{D}^{-1}$ and expand the fermion determinant,
\begin{equation}
\operatorname{Tr}\ln \left[\bar{D}^{-1}+ \delta \sigma (\mathbf{x})\right] = \operatorname{Tr}\ln (\bar{D}^{-1}) - \sum_{n = 1}^{\infty}\frac{1}{n}\operatorname{Tr}\left(-\bar{D}\delta \sigma (\mathbf{x})\right)^{n}.
\end{equation}
Consequently, the full effective action is expanded as a perturbative series in powers of $\delta\sigma$,
\begin{equation}
\mathcal{S}_{\mathrm{eff}} = \sum_{n = 0}^{\infty}\mathcal{S}_{\mathrm{eff}}^{(n)},
\end{equation}
where $\mathcal{S}_{\mathrm{eff}}^{(n)}$ collects all $n$-th order contributions. The three lowest-order terms are
\begin{equation}
\mathcal{S}_{\mathrm{eff}}^{(0)} = \beta V\frac{\bar{\sigma}^2}{4G} -\mathrm{Tr}\ln (\bar{D}^{-1}), 
\end{equation}
\begin{equation}
\mathcal{S}_{\mathrm{eff}}^{(1)} = \int d^3\mathbf{x}  \beta \frac{\bar{\sigma}\delta\sigma(\mathbf{x})}{2G}  - \mathrm{Tr}[\bar{D}\delta \sigma (\mathbf{x})], 
\end{equation}
\begin{equation}
\mathcal{S}_{\mathrm{eff}}^{(2)} = \int d^3\mathbf{x} \beta \frac{[\delta\sigma(\mathbf{x})]^2}{4G} + \mathrm{Tr}\frac{[\bar{D}\delta\sigma(\mathbf{x})]^2}{2}.\label{second_oder}
\end{equation}
The zeroth-order term yields the homogeneous effective potential $\beta V U(\bar{\sigma})$. The linear term $\mathcal{S}_{\mathrm{eff}}^{(1)}$ vanishes when $\bar{\sigma}$ satisfies the homogeneous gap equation. The quadratic term $\mathcal{S}_{\mathrm{eff}}^{(2)}$ determines the stability; a negative quadratic direction is sufficient to destabilize the stationary state. If a mode is marginal, higher-order terms are required to determine its behavior. Note that the expansion also cannot exclude a distant minimum separated by a free-energy barrier.

To define the inverse propagator in eq. (\ref{second_oder}) in momentum space, we introduce the Fourier transform of the static scalar fluctuation,
\begin{equation}
\delta \tilde{\sigma} (\mathbf{p}) = \int d^3\mathbf{x} \, e^{-i\mathbf{p}\cdot \mathbf{x}}\delta \sigma (\mathbf{x}).
\end{equation}
Since $\delta \sigma (\mathbf{x})$ is real, the Fourier components satisfy $\delta \tilde{\sigma} (-\mathbf{p}) = \delta \tilde{\sigma}^{*}(\mathbf{p})$. With this convention, the quadratic effective action is written as
\begin{equation}
\mathcal{S}_{\mathrm{eff}}^{(2)} = \frac{\beta}{2}\int \frac{d^3\mathbf{p}}{(2\pi)^3} \Gamma^{(2)}(\bar{\sigma},\mu,B,T;\mathbf{p}) \delta \tilde{\sigma} (\mathbf{p})\delta \tilde{\sigma} (-\mathbf{p}),
\end{equation}
where the scalar inverse propagator (bosonic two-point function) is decomposed as
\begin{equation}
\Gamma^{(2)}(\bar{\sigma},\mu,B,T;\mathbf{p}) = \frac{1}{2G} + \Pi (\mathbf{p}),
\end{equation}
with the polarization function $\Pi(\mathbf{p})$ given by the quark loop,
\begin{equation}
\Pi (\mathbf{p}) = N_{c}\sum_{f = u,d}\frac{1}{\beta}\sum_{n}\int \frac{d^{3}\mathbf{k}}{(2\pi)^{3}}\mathrm{tr}\left[\tilde{D}_{f}(\omega_{n},\mathbf{k})\tilde{D}_{f}(\omega_{n},\mathbf{k} + \mathbf{p})\right],
\label{eq:Pi_formal}
\end{equation}
Here, $\tilde{D}_{f}(\omega_n, \mathbf{k})$ denotes the translationally invariant part of the quark propagator evaluated in momentum space. Note that because the external scalar perturbation is static (zero external energy), the energy flow is conserved in the fermion loop, and both propagators share the same Matsubara frequency $\omega_m$.

To justify the use of the translationally invariant part $\tilde{D}_{f}$, we note that in coordinate space, the full Dirac propagator $\bar{D}_{f}(x,y)$ contains a gauge-dependent Schwinger phase, i.e.,
\begin{equation}
\bar{D}_{f}(x,y) = e^{i\Phi_{f}(x,y)}\tilde{D}_{f}(x - y),
\label{eq:Schwinger_phase}
\end{equation}
where $\tilde{D}_{f}(x - y)$ denotes the translationally invariant part. For the neutral scalar channel considered here, the two propagators in the closed fermion loop carry opposite Schwinger phases. Because the phase is antisymmetric, $\Phi_{f}(y,x) = -\Phi_{f}(x,y)$, their product is unity,
\begin{equation}
e^{i\left[\Phi_{f}(x,y)+\Phi_{f}(y,x)\right]} = 1.
\end{equation}
Therefore, the Schwinger phases exactly cancel in the loop, and the polarization function can be evaluated entirely in terms of $\tilde{D}_{f}(\omega_n, \mathbf{k})$ in momentum space. This gauge invariance greatly simplifies the subsequent calculations.

The explicit Dirac trace, transverse Landau-level integrals, and Matsubara summation are presented in Appendix~\ref{app:Gamma2_derivation}. The resulting two-point function is
\begin{align}
\Gamma^{(2)}({\bf p}_\perp,p_z) &= \frac{1}{2G} + \sum_f \frac{N_c |q_f|B}{2\pi^2} e^{-x_f} \sum_{l,l'=0}^{\infty} x_f^{|l-l'|} \int_{-\infty}^{\infty}dk_z \bigg\{ -A_{f,l}(k_z)J_f(l,l';{\bf p}_\perp) \nonumber \\
&\quad + \frac{ \left( p_z^2+p_zk_z+2|q_f|B l'+2\bar{\sigma}^2 \right) J_f(l,l';{\bf p}_\perp) - K_f(l,l';{\bf p}_\perp) }{ 2p_zk_z+p_z^2+2|q_f|B(l'-l) } \left[ A_{f,l}(k_z)-A_{f,l'}(k_z+p_z) \right] \bigg\},
\label{eq:B_Gamma2_final}
\end{align}
where $x_f={\bf p}_\perp^2/(2|q_f|B)$.  Here, $l$ and $l'$ denote the Landau-level indices of the two quarks in the fermion loop, and we have defined
\begin{equation}
A_{f,l}(k_z) = \frac{ 1-\eta(E_{f,l}(k_z))-\bar{\eta}(E_{f,l}(k_z)) }{ 2E_{f,l}(k_z) },
\label{eq:B_Afn}
\end{equation}
and
\begin{align}
J_f (l,l';{\bf p}_\perp)= \frac{l_<!}{l_>!}\left[L_{l_<}^{|l-l'|}(x_f)\right]^2 + \frac{(l_< - 1)!}{(l_> - 1)!} \left [L_{l_< - 1}^{|l-l'|}(x_f)\right]^2, 
\end{align}
\begin{align}
K_f(l,l';{\bf p}_\perp) = \frac{4|q_f|Bl_<!}{(l_> - 1)!} L_{l_<}^{|l-l'|}(x_f)L_{l_< - 1}^{|l-l'|}(x_f),
\end{align}
with $l_<=\min(l,l'),\ l_>=\max(l,l')$. Here, $L_l^\alpha(x)$ denote the generalized Laguerre polynomials, which arise from the transverse part of the Landau-level wave functions. Terms containing a negative factorial or a negative Laguerre index (i.e., when $l_< = 0$ and $l_< - 1 < 0$) are understood to vanish.

In the loop contribution to Eq.~\eqref{eq:B_Gamma2_final}, the coupling between different Landau levels is governed by the transverse overlap factors $J_f$ and $K_f$. Physically, these factors represent the spatial overlap integrals of the Landau-level wave functions for the initial and final quark states. They vanish when the levels are orthogonal and are exponentially suppressed when the transverse momentum transfer ${\bf p}_\perp$ is large, due to the overall Gaussian factor $e^{-x_f} = \exp\left[-{\bf p}_\perp^2/(2|q_f|B)\right]$. Because of this exponential suppression, the dominant physical contributions are strongly localized in the region of small transverse momentum. Therefore, to capture the leading instability, it is both physically justified and analytically convenient to first consider the limit ${\bf p}_\perp = 0$ (i.e., $x_f = 0$). In this limit, the Gaussian factor becomes unity, and the term $x_f^{|l-l'|}$ eliminates all off-diagonal contributions ($l \neq l'$), leaving only the diagonal Landau-level terms ($l'=l$).  Using $L_l(0)=1$, the surviving transverse factors reduce to $J_f=\alpha_l$ and $K_f=4|q_f|Bl$, where $\alpha_l=2-\delta_{l0}$. Since $|q_f|Bl(\alpha_l - 2) = 0$, the transverse terms in the numerator cancel, and the two-point function becomes
\begin{align}
\Gamma^{(2)}(0,p_z) = \frac{1}{2G} +\sum_{f}\frac{N_c|q_f|B}{2\pi^2} \sum_{l=0}^{\infty}\alpha_l \int_{-\infty}^{\infty}dk_z \left\{ -A_{f,l}(k_z)  + \frac{p_z^2+p_zk_z+2\bar{\sigma}^2} {p_z(2k_z+p_z)} \left[A_{f,l}(k_z)-A_{f,l}(k_z+p_z)\right] \right\}.
\label{eq:Gamma2_longitudinal_main}
\end{align}
The Landau-level dependence remains in $E_{f,l}(k_z)$ and $A_{f,l}(k_z)$, so this reduction retains all levels. 

Even though the integral in Eq.~\eqref{eq:Gamma2_longitudinal_main} is divergent and requires regularization, the zero-momentum limit obeys the homogeneous curvature identity, provided that the regulator is independent of $\bar{\sigma}$,
\begin{equation}
\Gamma^{(2)}(0,0) = \frac{\partial^2 U(\bar{\sigma},\mu,T,B)}{\partial\bar{\sigma}^2},
\label{eq:homogeneous_curvature_identity}
\end{equation}
whenever this curvature exists. The limit is taken on the combined occupation difference in Eq.~\eqref{eq:Gamma2_longitudinal_main}, before treating its apparently singular factors separately. This identity provides a robust analytical and numerical consistency check. It ensures that the zero-momentum limit of our full momentum-dependent loop calculation exactly recovers the thermodynamic curvature mass obtained from the effective potential.

Having established the formal expression for the bosonic two-point function, we now clarify its physical interpretation as a stability criterion. If $\Gamma^{(2)}({\bf p}) > 0$ for all momenta, the quadratic form is positive definite, and the homogeneous state is stable in the scalar sector considered here. Conversely, if $\Gamma^{(2)}({\bf p}) < 0$ at some nonzero momentum, an infinitesimal scalar modulation lowers the free energy, establishing an instability of the homogeneous reference state. It should be noted that while a negative $\Gamma^{(2)}({\bf p})$ establishes a strong tendency towards an inhomogeneous phase, the specific equilibrium profile and modulation amplitude can only be determined by minimizing the full nonlinear effective action.

%%%%%%%%%%%%%%%%%%%%%%%%%%%%%%%%%%%%%
\section{Momentum-Dependent Structures of the Bosonic Two-Point Function}
%%%%%%%%%%%%%%%%%%%%%%%%%%%%%%%%%%%%%
\label{sec:numerical_analysis}

In this section, we first analytically investigate the longitudinal slice (${\bf p}_\perp = 0$) of the two-point function at zero temperature to extract the leading singularities. Subsequently, we perform a comprehensive numerical evaluation of the full momentum-space topography, analyzing its thermal evolution and the emergence of a moat regime. To circumvent the limitations of the stability analysis in the broken phase, we evaluate the two-point function exclusively in the symmetry restored phase ($\bar{\sigma} = 0$). We select two fixed chemical potentials, $\mu = 350$ MeV and $\mu = 450$ MeV, ensuring that the homogeneous background remains chirally symmetric across all the magnetic field values considered below. This setup establishes a common symmetric reference state, enabling a systematic comparison of how spatially modulated scalar fluctuations respond to varying magnetic fields.

%%%%%%%%%%%%%%%%%%%%%%%%%%%%%%%%%%%%%%%%%%%%%%%%%%%%%%%%%
\subsection{Logarithmic singularity at zero temperature and Peierls' instability}
%%%%%%%%%%%%%%%%%%%%%%%%%%%%%%%%%%%%%%%%%%%%%%%%%%%%%%%%%
\label{subsec:peierls}

In the following, we will first focus on the longitudinal slice ${\bf p}_\perp = 0$, where only the diagonal Landau-level terms ($l'=l$) contribute to the two-point function. We observe that the denominator of the second term in eq.~\eqref{eq:Gamma2_longitudinal_main} develops an apparent singularity at $k_z = -p_z/2$.

It is instructive to contrast this with the non-magnetized cases. In a genuine $(1+1)$-dimensional system (such as the Gross-Neveu model), this singularity manifests as a severe infrared divergence. In the $(3+1)$-dimensional NJL model without a magnetic field, however, this singularity is innocuous: the continuous integration over the transverse momentum phase space effectively smooths it out, yielding a finite result. In the present magnetized system, the situation is fundamentally different. Landau quantization freezes the transverse motion and effectively reduces the system to a collection of $(1+1)$-dimensional subsystems. As we will demonstrate below, this dimensional reduction prevents the singularity from being smoothed out, allowing it to survive as a true logarithmic divergence at zero temperature.

At $T = 0$, the Fermi-Dirac distributions reduce to step functions, and the coefficient $A_{f,l}(k_z)$ simplifies to
\begin{equation}
A_{f,l}(k_z) = \frac{\Theta(E_{f,l}(k_z) - \mu)}{2E_{f,l}(k_z)}.
\end{equation}
The relevant physical excitation is governed by the difference $A_{f,l}(k_z) - A_{f,l}(k_z + p_z)$, which dictates the creation of particle-hole pairs that can bridge states near the Fermi surface.

For a given Landau level and flavor, the Fermi momentum $k_{f,l}^F$ is defined by the positive root of $E_{f,l}(k_{f,l}^F) = \mu$. Letting $E \equiv E_{f,l}(k_z)$ and $E' \equiv E_{f,l}(k_z + p_z)$, the denominator of the second term in Eq.~\eqref{eq:Gamma2_longitudinal_main} becomes $p_z(2k_z + p_z) = E'^2 - E^2$, which has an apparent singularity at $k_z = -p_z/2$. However, this singularity is removable as long as $p_z \neq 2k_{f,l}^F$.
For $0 < p_z < 2k_{f,l}^F$, both energies at the singular point are below the Fermi surface ($E, E' < \mu$). The step functions vanish, so the numerator is identically zero. For \(p_z > 2k_{f,l}^F\), both energies are above the Fermi surface ($E, E' > \mu$). The step functions equal unity, and the difference $1/(2E) - 1/(2E')$ is proportional to $E' - E$, yielding a finite value
\begin{equation}
\frac{1}{2EE'(E + E')}.
\end{equation}

Only at the critical momentum $p_z = 2k_{f,l}^F$ does a true non-removable singularity emerge. At this point, we have $E = E' = \mu$ at $k_z = -k_{f,l}^F$. To be precise, let $k_z = -k_{f,l}^F + \delta$ with an infinitesimal offset $\delta$. For $\delta \to 0^-$, we have $|k_z| > k_{f,l}^F$ and $|k_z + p_z| < k_{f,l}^F$, yielding $E_{f,l}(k_z) > \mu$ and $E_{f,l}(k_z + p_z) < \mu$. The difference $A_{f,l}(k_z) - A_{f,l}(k_z + p_z)$ approaches $+1/(2\mu)$. Conversely, for $\delta \to 0^+$, the two momenta cross the Fermi surface in opposite directions, $|k_z| < k_{f,l}^F$ and $|k_z + p_z| > k_{f,l}^F$, and the difference approaches $-1/(2\mu)$. Therefore, the integrand exhibits the asymptotic behavior
\begin{equation}
\lim_{p_z \to 2k_{f,l}^F}\frac{A_{f,l}(k_z) - A_{f,l}(k_z + p_z)}{p_z(2k_z + p_z)} \sim -\frac{1}{8\mu k_{f,l}^F|k_z + k_{f,l}^F|}.
\end{equation}
Integrating this singular term over $k_z$ yields a logarithmic divergence. Consequently, For a fixed flavor $f$ and Landau level $l$, the leading singular contribution to the two-point function is given by
\begin{equation}
\delta\Gamma_{f,l}^{(2)}(0,p_z) = \frac{N_c|q_f|B\alpha_lk^F_{f,l}}{4\pi^2 \mu}\ln \left|\frac{p_z - 2k_{f,l}^F}{p_z + 2k_{f,l}^F}\right|.
\end{equation}
Since its coefficient is positive, $\delta\Gamma^{(2)}(0,p_z) \to -\infty$ as $p_z \to 2k_{f,l}^F$. Contributions at the same singular momentum add. In particular, $q_u = 2q_d$ implies $k_{u,l}^F = k_{d,2l}^F$, so these two flavor contributions reinforce the same logarithmic divergence.

 In a 1D system, the Fermi surface consists of just two discrete points, $\pm k_{f,l}^F$. A scalar perturbation with momentum $p_z = 2k_{f,l}^F$ perfectly connects these opposite Fermi points, allowing a particle at $+k_{f,l}^F$ to scatter into an empty state at $-k_{f,l}^F$ (and vice versa). This phenomenon, known as Fermi surface nesting, produces a divergent number of low-energy particle-hole excitations. Consequently, the polarization loop exhibits a negative logarithmic divergence, driving $\Gamma^{(2)}(0,p_z)$ to negative infinity. This is the well-known Peierls mechanism \cite{Peierls1930, Frohlich1954, Pouget2016}. Importantly, the sum over Landau levels does not cancel this divergence. Each occupied Landau level has a distinct Fermi momentum $k_{f,l}^F$ and independently contributes a logarithmic singularity at its corresponding momentum $p_z = 2k_{f,l}^F$. These singularities occur at different locations in momentum space, and they can even reinforce each other when $k_{f,l}^F = k_{f',l'}^F$ (e.g., between the $l$-th level of the $u$-quark and the $2l$-th level of the $d$-quark). Physically, this implies that the homogeneous state is unstable: by spontaneously developing a spatially modulated condensate with wave vector $ 2k_{f,l}^F$, the system can open a gap at the Fermi surface and lower its total free energy. Since the lowest Landau level (LLL) has $k_{f,0}^F = \mu$, it guarantees such negative values of $\Gamma^{(2)}$ near $p_z = 2\mu$ for any $B > 0$ at $T = 0$. Related logarithmic divergences also occur in the magnetized electron-gas response \cite{Peierls_in_electron_gas}.

It is worth noting that introducing a small explicit bare quark mass \(m \neq 0\) in the symmetric phase would slightly shift the nesting singularity from \(p_z = 2\mu\) to \(p_z = 2\sqrt{\mu^2 - m^2}\) and slightly reduce the coefficient of the logarithmic divergence. Physically, the bare mass explicitly breaks chiral symmetry, opening a mass gap near the Fermi surface that increases the energy cost for particle-hole excitations, thereby weakening the pairing tendency. However, the qualitative feature of the Peierls instability driven by one-dimensional nesting remains robust. 

At ${\bf p}_\perp \neq 0$, the transverse factors in eq.~\eqref{eq:B_Gamma2_final} must be retained. For $l' = l$, the singular momentum remains $p_z = 2k_{f,l}^F$, while its coefficient acquires the Laguerre factors. For two different levels with real positive Fermi momenta, the opposite Fermi points are connected at
\begin{equation}
p_z = k_{f,l}^F + k_{f,l'}^F, \qquad l \neq l'.
\end{equation}
At this momentum and $k_z = -k_{f,l}^F$, both energies equal $\mu$. The scalar numerator in eq.~\eqref{eq:B_Gamma2_final} becomes $(\mu^2 + k_{f,l}^Fk_{f,l'}^F)J_f - K_f$. The same integration across the two energy thresholds gives, for an ordered pair $(l,l')$ in the original double sum,
\begin{equation}
\delta\Gamma_{f,l,l'}^{(2)}(p_\perp,p_z) = \frac{N_c|q_f|B}{4\pi^2\mu(k_{f,l}^F + k_{f,l'}^F)} e^{-x_f}x_f^{|l - l'|}\left[(\mu^2 +k_{f,l}^Fk_{f,l'}^F)J_f - K_f\right]\ln \left|\frac{p_z - k_{f,l}^F - k_{f,l'}^F}{p_z + k_{f,l}^F + k_{f,l'}^F}\right|,\label{log_divergence_full_momenta}
\end{equation}
where 
\begin{equation}
|K_f| = 4|q_f|B\sqrt{l l'} \sqrt{\frac{l_< !}{l_> !}\frac{(l_< - 1)!}{(l_> - 1)!}}\left|L_{l_<}^{|l - l'|}(x_f)L_{l_< - 1}^{|l - l'|}(x_f)\right| \leq 2|q_f|B\sqrt{l l'} J_f.
\end{equation}
We thus have
\begin{equation}
(\mu^2 +k_{f,l}^Fk_{f,l'}^F)J_f - K_f \geq \left[\mu^2 +k_{f,l}^Fk_{f,l'}^F - 2|q_f|B\sqrt{l l'}\right]J_f.
\end{equation}
Since \(\mu^2 > 2|q_f|B\sqrt{l l'}\), the expression in square brackets is strictly positive. Therefore, the logarithmic coefficient in eq. (\ref{log_divergence_full_momenta}) is positive whenever the transverse overlap is nonzero. A similar logarithmic divergence has been reported in the study of the magnetic dual chiral density wave (MDCDW) phase \cite{Gyory2022}. Specifically, the free-energy curvature with respect to the condensate magnitude diverges at a particular modulation momentum. This divergence shares the same physical origin as our result: it is driven by one-dimensional Fermi surface nesting, where particle-hole excitations perfectly connect the opposite Fermi points of the Landau levels. In the context of MDCDW, this divergent curvature prevents the system from reaching the chirally symmetric phase, giving rise to a small but non-zero ``remnant mass'' over a wide range of chemical potentials. In our stability analysis, the corresponding logarithmic divergence in the two-point function directly signals the absolute instability of the homogeneous state against spatial modulation. While these observations establish a strong tendency towards crystallization, determining the exact equilibrium profile and modulation amplitude requires a full nonlinear minimization of the thermodynamic potential.

The crucial role of the magnetic field in triggering this instability becomes manifest by examining the zero-field limit. For any $B \neq 0$, Landau quantization effectively reduces the system to a collection of 1+1-dimensional subsystems. Each occupied Landau level independently yields a sharp logarithmic singularity at its corresponding Fermi momentum. It is worth noting that the strength (or weight) of each divergence is proportional to $|q_f|B$. Therefore, in the weak-field limit, the individual singularities become weaker, while their spacing in momentum space shrinks ($\Delta k_{l}^F \sim B$). This high density of overlapping divergences indicates a strong instability of the homogeneous state. However, only in the strict $B = 0$ limit does the discrete sum over Landau levels transition into a continuous integral over the transverse momentum ${\bf k}_\perp^2$. This restores the full three-dimensional phase space, completely smoothing out the singularities. Consequently, the two-point function remains finite for all momenta, including the would-be critical momentum $p = 2\mu$. This contrast highlights that the severe Peierls instability is a genuine consequence of Landau quantization and the resulting dimensional reduction along the magnetic field axis.

%To verify this, we first take the \(B \to 0\) limit at fixed \(T > 0\) and then take \(T %\to 0\). At \(T = 0\), the complete zero-field two-point function with the same cutoff, %expressed in terms of \(p = |\mathbf{p}|\), is
%\begin{equation}
%\Gamma_{B = 0}^{(2)}(p) = \frac{1}{2G} +4N_{c}N_{f}\int %\frac{d^{3}k}{(2\pi)^{3}}\left\{-A_{0}(k) + \frac{p^{2} + %\mathbf{k}\cdot\mathbf{p}}{p^{2} + 2\mathbf{k}\cdot\mathbf{p}}\times \left[A_{0}(k) %- A_{0}(|\mathbf{k} + \mathbf{p}|)\right]\right\},
%\label{eq:Gamma2_B0}
%\end{equation}
%where \(k = |\mathbf{k}|\), \(N_f = 2\), and \(A_0(k) = \Theta(k - \mu)/(2k)\). The %smooth and finite profile of Eq.~\eqref{eq:Gamma2_B0} provides a baseline, %demonstrating that without the dimensional reduction induced by the magnetic field, %the homogeneous state remains stable against scalar modulations.

At finite temperature, the Fermi-Dirac distribution changes continuously across the Fermi energy. The sharp step functions in $A_{f,l}$ are replaced by smooth thermal distributions, which significantly alters the behavior of the integrand near the singularity $p_z = 2k_{f,l}^F$. 
Physically, thermal fluctuations smear the sharp Fermi surface over an energy range of order \(T\). Consequently, the strict momentum matching condition is relaxed, and particle-hole excitations can only occur within a thermal shell of width \(\Delta k_z \sim T/v_{f,l}^F\) around the Fermi surface, where \(v_{f,l}^F = k_{f,l}^F/\mu\) is the Fermi velocity for the \(l\)-th Landau level. 

Focusing on the longitudinal slice \(p_\perp = 0\), we rigorously demonstrate how the thermal width regulates the pole by introducing the variable \(x \equiv k_z + k_{f,l}^F\) and evaluating the integrand near the singularity point \(p_z = 2k_{f,l}^F\). The energies of the initial and final states are linearized as
\begin{equation}
E_{f,l}(k_z) \simeq \mu - v_{f,l}^F x, \qquad E_{f,l}(k_z + p_z) \simeq \mu + v_{f,l}^F x.
\end{equation}
The relevant difference of the thermal distribution functions then takes the form
\begin{equation}
A_{f,l}(k_z) - A_{f,l}(k_z + p_z) \sim \tanh\left(\frac{v_{f,l}^F x}{2T}\right).
\end{equation}
Meanwhile, the denominator of the singular term simplifies to \(p_z(2k_z + p_z) \simeq 4k_{f,l}^F x\). Consequently, the behavior of the integrand near the pole is governed by
\begin{equation}
\frac{A_{f,l}(k_z) - A_{f,l}(k_z + p_z)}{p_z(2k_z + p_z)} \sim \frac{1}{x} \tanh\left(\frac{v_{f,l}^F x}{2T}\right).
\end{equation}
The thermal distribution regulates this singularity in two distinct regimes. Far from the singularity (\(|x| \gg T/v_{f,l}^F\)), the hyperbolic tangent approaches unity, \(\tanh(v_{f,l}^F x / 2T) \to 1\), and the integrand reduces to \(\sim 1/x\), yielding the expected logarithmic behavior. Conversely, in the immediate vicinity of the singularity (\(|x| \ll T/v_{f,l}^F\)), the hyperbolic tangent is linearized as \(\tanh(v_{f,l}^F x / 2T) \simeq v_{f,l}^F x / 2T\), which cancels the \(1/x\) pole and renders the integrand finite and regular as \(x \to 0\). This mechanism effectively regularizes the logarithmic integral at the infrared scale \(|x| \sim T/v_{f,l}^F\). Consequently, instead of diverging to negative infinity as in the zero-temperature case, the integral yields a finite negative value, which scales as \(\ln(T/v_{f,l}^F)\).  Therefore, the logarithmic divergences associated with the sharp zero-temperature Fermi surface are removed, and \(\Gamma^{(2)}\) becomes finite at the corresponding momenta. The resulting finite negative minimum acts as a precursor to the instability, with its depth depending on the temperature.

%%%%%%%%%%%%%%%%%%%%%%%%%%%%%%%%%%%%%%%%%%%%%%%%%%%%%%%%%
\subsection{Numerical analysis and the momentum-space topography}
%%%%%%%%%%%%%%%%%%%%%%%%%%%%%%%%%%%%%%%%%%%%%%%%%%%%%%%%%
\label{subsec:num_results}

In this subsection, we present the numerical evaluation of the full momentum-dependent  two-point function $\Gamma^{(2)}({\bf p}_{\perp}, p_{z})$ at two different chemical potentials, $\mu = 350$ MeV and $450$ MeV, focusing on the symmetric restored phase ($\bar{\sigma} = 0$). Since the analytical singularity structure at zero temperature has been thoroughly established in Sec.~\ref{subsec:peierls}, here we focus directly on the thermal evolution at different temperatures. We investigate how the finite-temperature effects and varying magnetic fields reshape the momentum-space topography of $\Gamma^{(2)}$, ranging from deep instability canyons to stable moat precursors.

\begin{figure*}[t]
	\centering
	\subfloat[]{%
		\begin{minipage}[b]{0.45\textwidth}
			\centering
			\includegraphics[width=\linewidth]{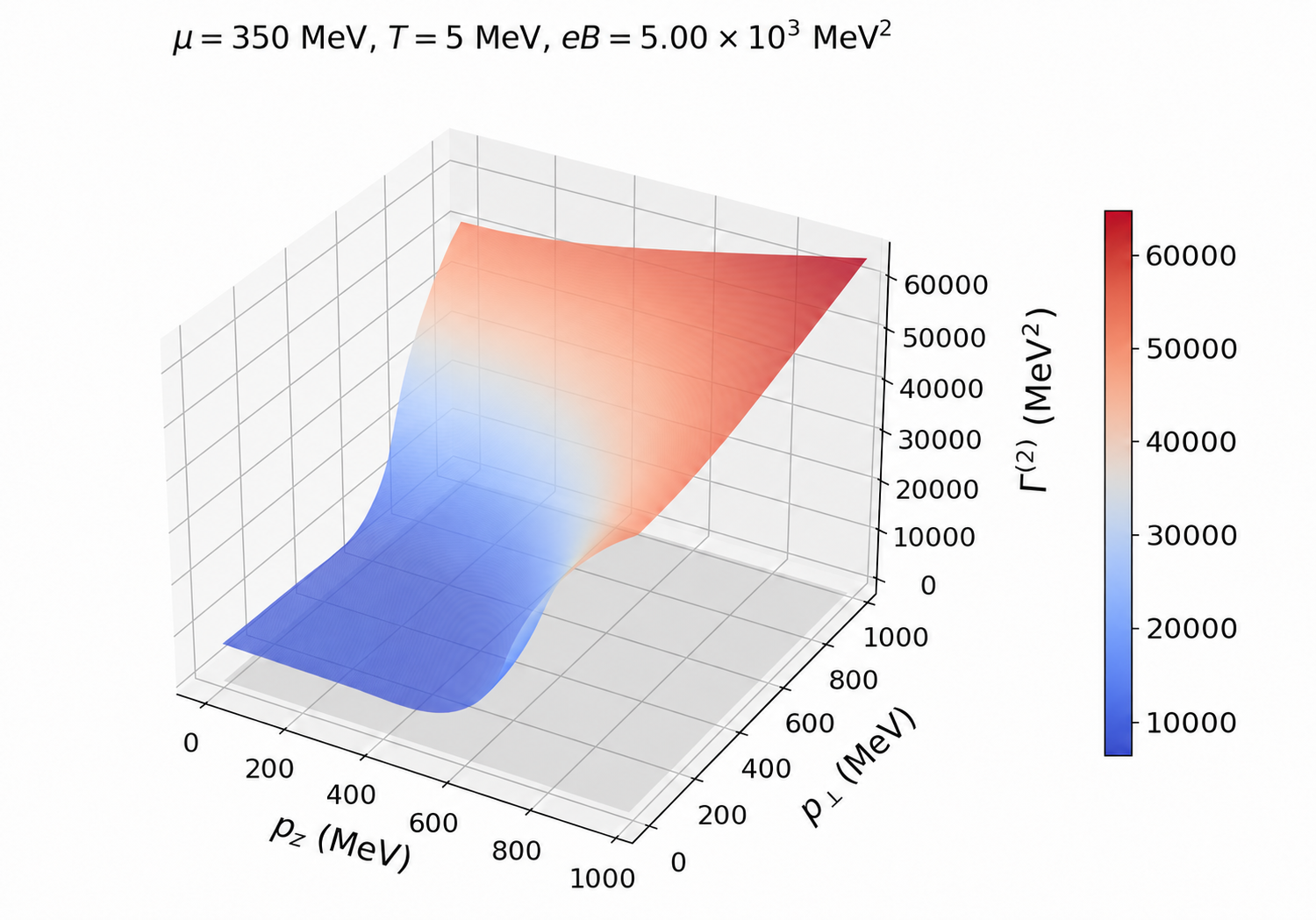}
		\end{minipage}%
	}
	\hspace{2cm}
	\subfloat[]{%
		\begin{minipage}[b]{0.4\textwidth}
			\centering
			\includegraphics[width=\linewidth]{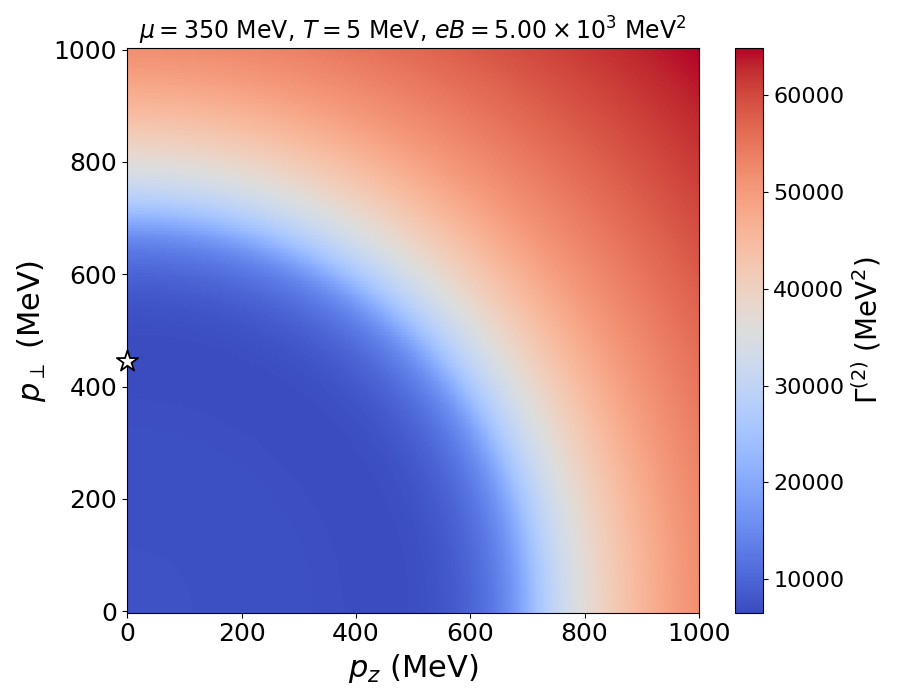}
		\end{minipage}%
	}
	\caption{Static two-point function $\Gamma^{(2)}$ at a weak magnetic field  $eB = 5 \times 10^3$ MeV$^2$, finite temperature $T=5$ MeV, and chemical potential $\mu = 350$ MeV. (a) Three-dimensional surface. (b) Color map. The star marks global minimum of \(\Gamma^{(2)}\) on the numerical grid.}
	\label{fig:fig2}
\end{figure*}

\begin{figure*}[t]
	\centering
	\subfloat[]{%
		\begin{minipage}[b]{0.45\textwidth}
			\centering
			\includegraphics[width=\linewidth]{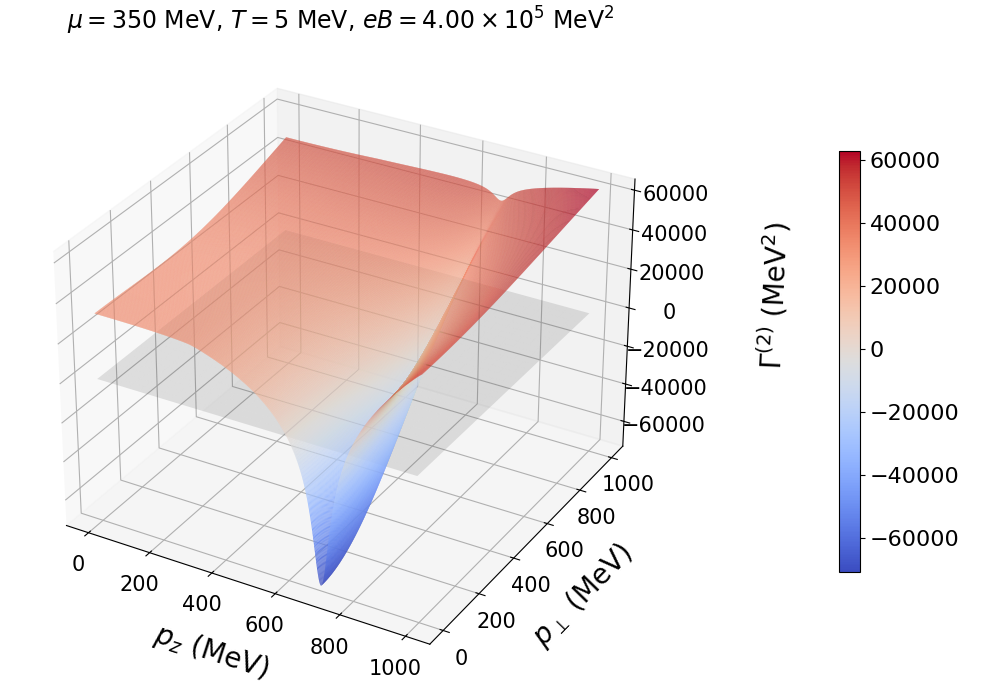}
		\end{minipage}%
	}
	\hspace{2cm}
	\subfloat[]{%
		\begin{minipage}[b]{0.4\textwidth}
			\centering
			\includegraphics[width=\linewidth]{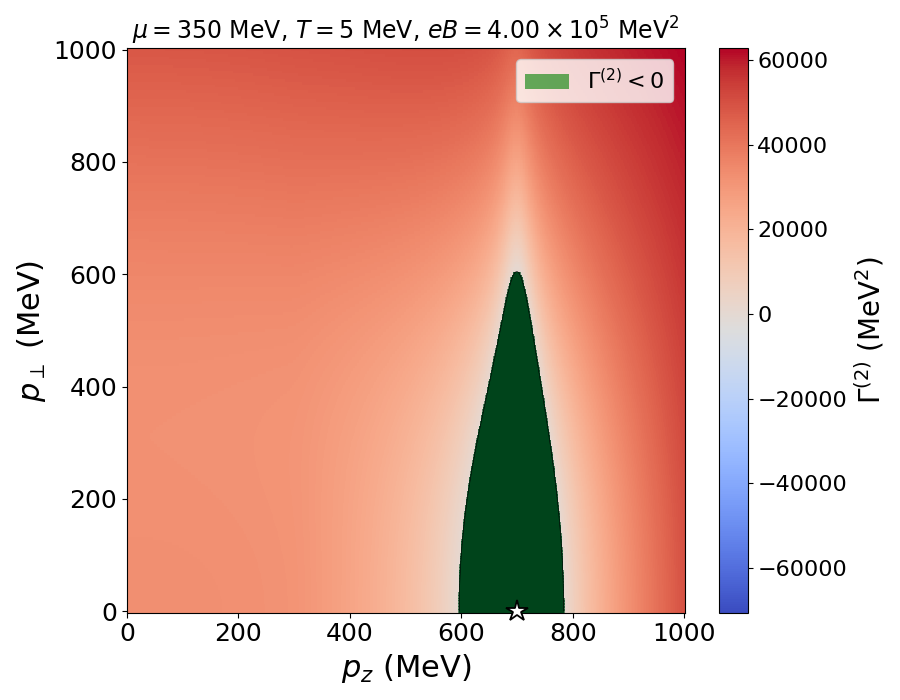}
		\end{minipage}%
	}
	\caption{ Same as Fig.~\ref{fig:fig2}, but for $eB = 4 \times 10^5$ MeV$^2$. The green shaded area in the color map indicates the region where $\Gamma^{(2)} < 0$,  and the star marks global minimum of $\Gamma^{(2)}$ on the numerical grid.}
	\label{fig:fig3}
\end{figure*}

As a starting point, we consider the case of a relatively weak magnetic field at a finite temperature, as shown in Fig.~\ref{fig:fig2} ($eB = 5 \times 10^3$ MeV$^2$, $T = 5$ MeV, and $\mu = 350$ MeV). The numerical surface remains strictly positive definite across the entire momentum space. This is because, in the weak-field limit, the Landau levels are densely packed, and the logarithmic divergences at zero temperature carry weights proportional to \(B\). Consequently, even a relatively small thermal broadening at \(T=5\) MeV is sufficient to completely smear out these weak divergences, lifting the negative regions into a shallow, smooth valley. 

It is particularly noteworthy that the global minimum of $\Gamma^{(2)}$ in this regime is located at $p_z = 0$ but at a non-zero transverse momentum ${\bf p}_\perp$. This behavior originates from the interplay of the transverse overlap factors. At exactly ${\bf p}_\perp = 0$, the orthogonality factor $x_f^{|l-l'|}$ strictly forbids off-diagonal transitions ($l \neq l'$), leaving only the diagonal contributions. As ${\bf p}_\perp$ increases from zero, the off-diagonal transitions are activated, and their collective contributions provide additional negative weights that lower the value of $\Gamma^{(2)}$. However, with a further increase of ${\bf p}_\perp$, the overall Gaussian suppression factor $e^{-p_\perp^2 / (2|q_f|B)}$ takes over, exponentially damping the integrand and pushing the function back up. The competition between the activation of off-diagonal transitions and the Gaussian suppression naturally generates a global minimum at a finite transverse momentum. Meanwhile, the weak $2k_F$ longitudinal nesting features are entirely masked by the thermal broadening at $T=5$ MeV, allowing the transverse dynamics to shape the global topography.

The situation changes dramatically when a strong magnetic field is applied. In Fig.~\ref{fig:fig3} ($eB = 4 \times 10^5$ MeV$^2$, $T = 5$ MeV, and $\mu = 350$ MeV), the Landau quantization becomes extremely pronounced. Since the magnetic field is so strong that only the lowest Landau level (LLL) crosses the Fermi surface, the dimensional reduction to a purely one-dimensional system becomes highly effective. Consequently, a deep and prominent canyon plunges into the negative region, visible as the green shaded area where $\Gamma^{(2)} < 0$. This robust negative minimum near $p_z \sim 700$ MeV clearly demonstrates the Peierls instability driven by the strong one-dimensional Fermi surface nesting at $p_z = 2k_{f,0}^F = 2\mu$.

It is worth noting that this deep negative valley occurs near $p_z \sim 700$ MeV, which is slightly above the ultraviolet cutoff $\Lambda = 653$ MeV adopted in our model. While values beyond $\Lambda$ lie outside the strict domain of validity of the NJL model as a low-energy effective theory, the emergence of this negative region demonstrates a robust kinematic feature of the one-dimensional nesting. In a more fundamental approach with running couplings, the exact location and depth of this valley would be modified by the asymptotic freedom of QCD, but the qualitative tendency towards spatial modulation remains a genuine physical effect.

\begin{figure*}[t]
	\centering
	\subfloat[]{%
		\begin{minipage}[b]{0.45\textwidth}
			\centering
			\includegraphics[width=\linewidth]{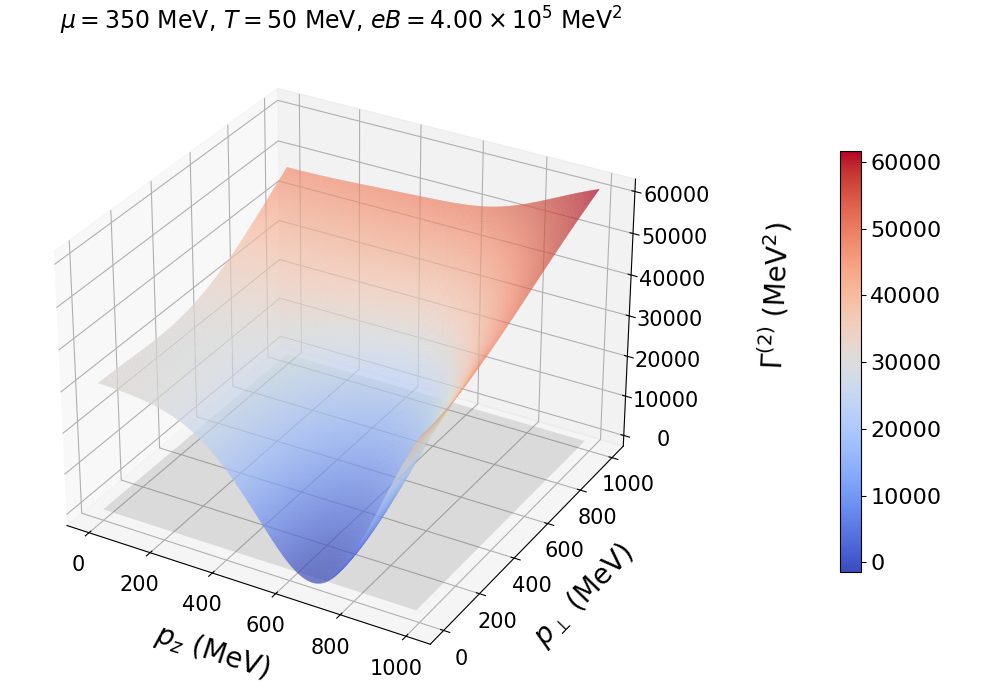}
		\end{minipage}%
	}
	\hspace{2cm}
	\subfloat[]{%
		\begin{minipage}[b]{0.4\textwidth}
			\centering
			\includegraphics[width=\linewidth]{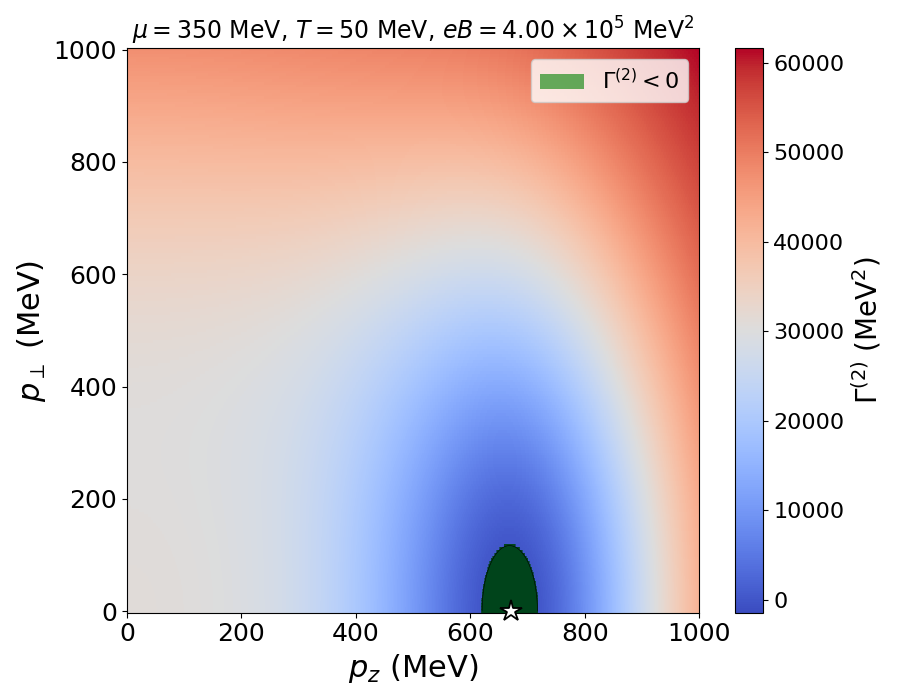}
		\end{minipage}%
	}
	\caption{Same as Fig.~\ref{fig:fig3}, but at a higher temperature $T = 50$ MeV. The deep negative canyon is lifted into a much shallower valley. The star marks the global minimum on the numerical grid.}
	\label{fig:fig4}
\end{figure*}

\begin{figure*}[t]
	\centering
	\subfloat[]{%
		\begin{minipage}[b]{0.45\textwidth}
			\centering
			\includegraphics[width=\linewidth]{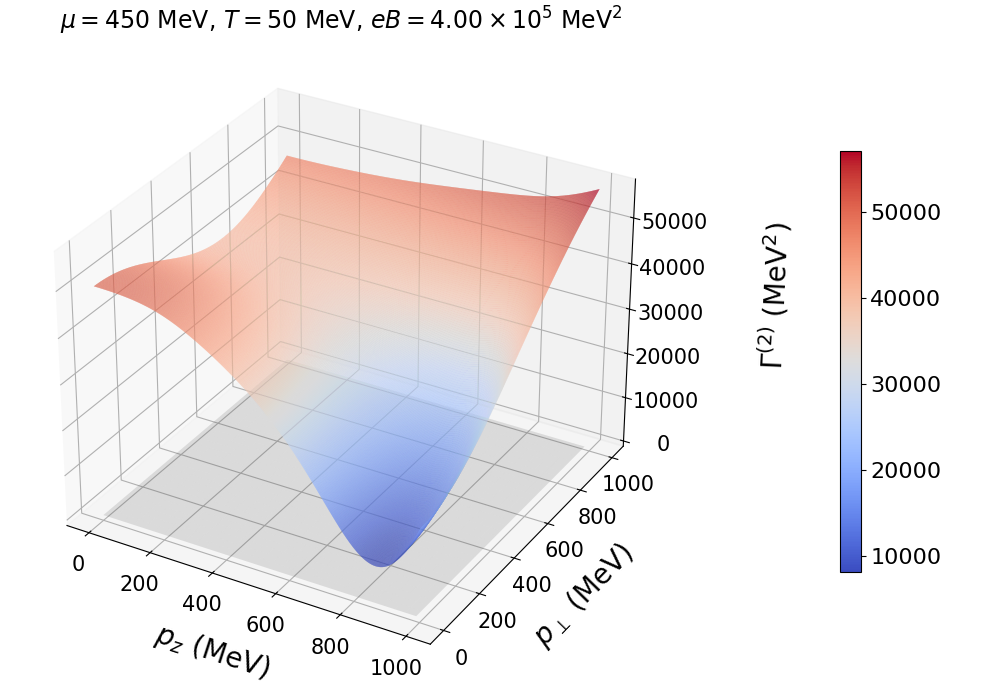}
		\end{minipage}%
	}
	\hspace{2cm}
	\subfloat[]{%
		\begin{minipage}[b]{0.4\textwidth}
			\centering
			\includegraphics[width=\linewidth]{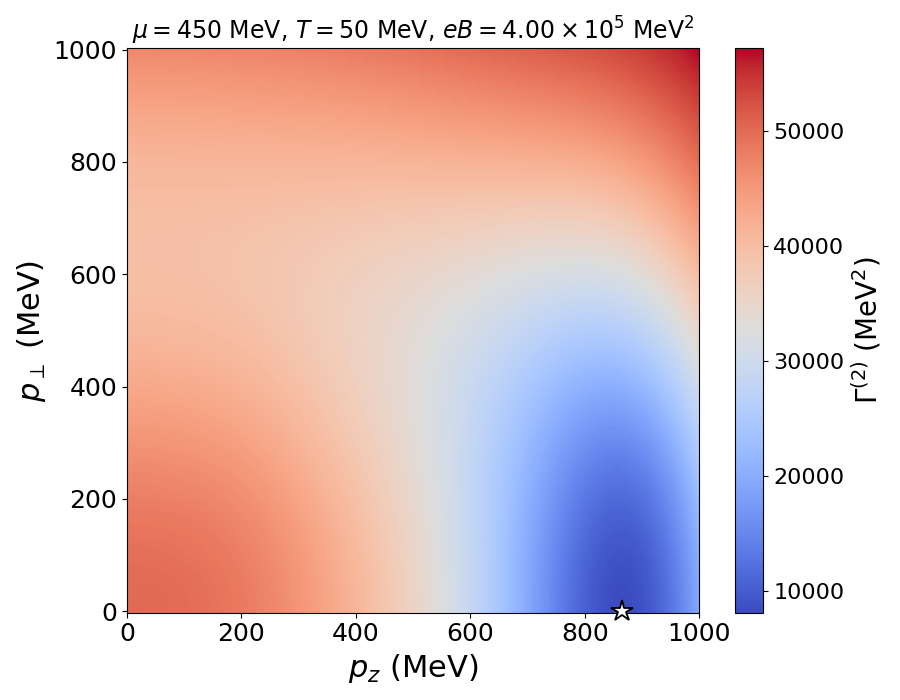}
		\end{minipage}%
	}
	\caption{Same as Fig.~\ref{fig:fig4}, but at a higher chemical potential $\mu = 450$ MeV. The two-point function is strictly positive. A broad positive basin persists at $p_z \sim 850$ MeV with a flat curvature at $p_z=0$. The star marks the global minimum on the numerical grid.}
	\label{fig:fig5}
\end{figure*}

Next, we examine the thermal evolution of this strong-field instability. Comparing Fig.~\ref{fig:fig3} with Fig.~\ref{fig:fig4},  we observe that increasing the temperature significantly alters the momentum topography. As thermal fluctuations smear the sharp Fermi surface, the strict momentum matching condition for particle-hole excitations is relaxed. The previously deep negative canyon is lifted and replaced by a much smoother and shallower valley. The negative region is almost entirely washed out, leaving only a minuscule negative remnant.  Furthermore, our numerical scan reveals that for $eB = 4 \times 10^5$ MeV$^2$, the negative regions completely disappear when the temperature reaches $T \sim 60$ MeV. This confirms that thermal broadening naturally provides an effective infrared cutoff for the logarithmic divergences, and a sufficiently high temperature can completely suppress the scalar instability associated with the zero-temperature Peierls mechanism.

Finally, Fig.~\ref{fig:fig5} shows the momentum topography at an even higher chemical potential (\(\mu = 450 \mathrm{MeV}\)) while maintaining \(T = 50 \mathrm{MeV}\) and \(eB = 4 \times 10^5 \mathrm{MeV}^2\). In this scenario, the two-point function is strictly positive everywhere, indicating that the homogeneous state is locally stable against inhomogeneous perturbations. However, it is particularly noteworthy that a broad, non-trivial positive basin persists around \(p_z \sim 850 \mathrm{MeV}\), while the curvature at the origin (\(p_z = 0\)) is relatively flat. This non-monotonic momentum dependence, where the global minimum is located at a non-zero momentum, is the hallmark of a \textit{moat regime} \cite{Pisarski2019, Pisarski2020}. It acts as a precursor state to the inhomogeneous phase, reflecting a strong residual tendency towards spatial modulation even though the absolute instability has been thermally melted.

To conclude, our numerical analysis reveals a rich interplay between Landau quantization and thermal fluctuations in the momentum space. The strong magnetic field promotes a one-dimensional Peierls instability at zero temperature, while finite temperature acts to melt this crystalline tendency, transitioning the system from an absolutely unstable canyon into a stable, yet highly susceptible, moat regime. This moat structure, characterized by a flat curvature at the origin and a non-trivial minimum at $p_z \simeq 2\mu$, suggests that the system sits on the verge of inhomogeneous chiral symmetry breaking. Although the strict stability at this parameter set is maintained, the presence of this soft mode could lead to observable consequences, such as modified meson dispersion relations and enhanced dilepton production in heavy-ion collisions \cite{Pisarski2021, Rennecke2023, Fukushima2024,Fu2025,Nussinov2024}. 

%Moreover, determining the actual equilibrium profile and amplitude of the potential %inhomogeneous phase would require a full nonlinear minimization beyond the %quadratic stability analysis adopted here, which remains an important direction for %future work.

%%%%%%%%%%%%%%%%%%%%%%%%%%%%%%%%%%%
\section{Conclusion and Outlook}
%%%%%%%%%%%%%%%%%%%%%%%%%%%%%%%%%%%
\label{sec:conclusion}

In this work, we have investigated the stability of homogeneous chiral condensates in magnetized quark matter using the two-flavor Nambu--Jona-Lasinio (NJL) model. By expanding the scalar-restricted effective action to quadratic order, we derived the full momentum-dependent bosonic two-point function $\Gamma^{(2)}({\bf p}_\perp, p_z)$ in the presence of a constant magnetic field. We then applied the stability analysis approach developed in \cite{Koenigstein2022} to test the homogeneous state against spatially modulated perturbations. Following the known validity criteria of this method, we focus our numerical evaluation on the symmetric restored phase ($\bar{\sigma} = 0$), where the quadratic expansion remains strictly reliable.

Our analytic results demonstrate that Landau quantization fundamentally alters the phase structure by effectively reducing the system to a collection of 1+1-dimensional subsystems. At zero temperature, each occupied Landau level produces a sharp logarithmic divergence at its corresponding Fermi momentum $p_z = 2k_{f,l}^F$. This strong one-dimensional Fermi surface nesting drives a severe Peierls instability, which is universally guaranteed by the lowest Landau level (LLL) near $p_z = 2\mu$ for any $B > 0$. In contrast, in the strict \(B=0\) limit, the continuous integration over the transverse momentum phase space restores three-dimensional dynamics and effectively regularizes these divergences into a finite result. This contrast confirms that the severe Peierls instability is a genuine consequence of magnetic-field-induced dimensional reduction.

Numerically, we mapped out the momentum-space topography of $\Gamma^{(2)}$ across different magnetic fields, temperatures, and chemical potentials. For a relatively weak magnetic field (e.g., $eB = 5 \times 10^3$ MeV$^2$), the divergences carry small weights proportional to $eB$, and a small temperature ($T=5$ MeV) is sufficient to smear them into a shallow valley, with the global minimum shifted to a non-zero transverse momentum due to the competition between off-diagonal transitions and Gaussian suppression. In contrast, for a strong magnetic field (e.g., $eB = 4 \times 10^5$ MeV${^2}$), the effective one-dimensional reduction becomes pronounced. We observed a deep negative canyon near $p_z \simeq 2\mu$, providing robust numerical evidence for the Peierls instability. Although this momentum slightly exceeds the ultraviolet cutoff $\Lambda$, it reflects a genuine kinematic feature of Fermi surface nesting. Furthermore, our finite-temperature analysis shows that thermal fluctuations naturally provide an infrared cutoff for these divergences. As the temperature increases, the negative canyon is lifted, eventually vanishing entirely around $T \sim 60$ MeV. At higher chemical potentials (e.g., $\mu = 450$ MeV), the system becomes strictly stable even for a strong magnetic field, yet a broad, non-trivial positive basin persists with a flat curvature at the origin. This non-monotonic dispersion is the hallmark of a moat regime, acting as a precursor to the inhomogeneous phase and signaling a strong residual tendency towards spatial modulation even when the absolute instability has been thermally melted.

It is worth emphasizing that the Peierls instability driven by the LLL is deeply connected to its non-trivial topological properties. As demonstrated in previous studies \cite{DCDW,Ferrer2018,Frolov2010,Ferrer2020,Efrain_review}, the spectral asymmetry of the LLL generates an anomalous term in the thermodynamic potential that favors the formation of the inhomogeneous condensate. More importantly, this topological effect alters the low-energy excitation spectrum, eliminating the soft transverse modes that usually plague one-dimensional modulations. While our stability analysis strictly establishes the onset of this instability in the homogeneous background, it is precisely this LLL-driven mechanism that guarantees the resulting magnetic dual chiral density wave (MDCDW) phase is highly resilient against the Landau-Peierls instability at finite temperature. Consequently, our stability analysis provides a microscopic insight into how the topological properties of the LLL dynamically trigger the instability tendency towards inhomogeneous phases in magnetized quark matter.

The findings presented here naturally open up several avenues for future research. A natural extension of our quadratic stability analysis is to determine the exact equilibrium profile and precise phase boundaries via a full nonlinear minimization of the effective action, since the current approach without assuming specific shapes only identifies the tendency toward spatial modulation, such nonlinear studies will be crucial to pin down the true ground state.  Furthermore, because the emergence of inhomogeneous phases in the NJL model is highly sensitive to the ultraviolet regularization scheme\cite{Pannullo2023b,Buballa2021}, it is essential to cross-check our predictions using alternative prescriptions, such as the magnetic-field-independent regularization \cite{Menezes2009, Avancini2019, Azeredo2606, BF_1, BF_2} or the renormalization group consistent regularization scheme \cite{Braun2019, Gholami2025}. 

To move towards a more realistic description of QCD, it is essential to go beyond the large-$N_c$ mean-field limit by incorporating bosonic quantum fluctuations, which will determine whether long-range crystalline order is disordered into a quantum liquid \cite{Pisarski2020}. Methodologically, recent developments utilizing the 2PI effective action \cite{Motta2023, Motta2024} and the test-function-free inhomogeneous chiral susceptibility \cite{Motta2024b} offer promising pathways to extend such stability analyses to more sophisticated truncations of magnetized QCD.  Finally, extending this framework to incorporate physical quark masses, running QCD couplings, and explicit magnetic-field-dependent running couplings \cite{Wen2016,Farias2016} would provide a more realistic mapping of the magnetized QCD phase diagram.

\begin{acknowledgments}

B.F. would like to thank E. J. Ferrer for valuable comments and discussions. This work was funded by the National Natural Science Foundation of China (NSFC), grant number 12075093.

\end{acknowledgments}

\appendix
\section{Detailed derivation of the Bosonic two-point function}
\label{app:Gamma2_derivation}

In this appendix, we present the intermediate steps leading from the formal polarization loop in eq.(\ref{eq:Pi_formal}) to the explicit expression for the scalar two-point function given in eq.(\ref{eq:B_Gamma2_final}). 

Starting from the polarization function,
\begin{equation}
\Pi (\mathbf{p}) = N_{c}\sum_{f }\frac{1}{\beta}\sum_{n}\int \frac{d^{3}\mathbf{k}}{(2\pi)^{3}}\mathrm{tr}\left[\tilde{D}_{f}(\omega_{n},\mathbf{k})\tilde{D}_{f}(\omega_{n},\mathbf{k} + \mathbf{p})\right], \label{eq:polarization_function}
\end{equation}
we insert the Landau-level representation of the translationally invariant fermion propagator. The propagator in momentum space is given by
\begin{equation}
\tilde{D}_f(\nu_n,\mathbf{k}) = \sum_{l = 0}^{\infty}\tilde{D}_{f,l}(\omega_n,\mathbf{k}),
\end{equation}
where
\begin{align}
\nonumber \tilde{D}_{f,l}(\omega_n,\mathbf{k}) = &-\frac{(-1)^le^{-\frac{k_\perp^2} {|q_f|B}} }{(i\omega_n + \mu)^2 - E_{f,l}^2(k_z)}\left\{ \left[\gamma^0(i\omega_n + \mu) - \gamma^3 k_z + \bar{\sigma}\right] \left[2\mathcal{P}_+^{(f)} L_l\left(\frac{2k_\perp^2}{|q_f|B}\right) - 2\mathcal{P}_-^{(f)} L_{l-1}\left(\frac{2k_\perp^2}{|q_f|B}\right)\right] \right.\\
&\left.+ 4(\gamma \cdot k)_\perp L_{l-1}^1\left(\frac{2k_\perp^2}{|q_f|B}\right) \right\},\label{eq:app_S_prop}
\end{align}
with  the energy \(E_{f,l}(k_z) = \sqrt{k_z^2 + 2Q_f l + \bar{\sigma}^2}\). Here, \(L_l(x)\) and \(L_l^1(x)\) are the generalized Laguerre polynomials, and \(\mathcal{P}_\pm^{(f)} = \frac{1}{2}(1 \pm i \mathrm{sgn}(q_fB) \gamma^1\gamma^2)\) are the spin projectors.

Substituting Eq.~\eqref{eq:app_S_prop} into Eq.~\eqref{eq:polarization_function}, we need to evaluate the Dirac trace and integrate over the transverse loop momentum \(\mathbf{k}_\perp\equiv (k_1,k_2)\). To facilitate this, we introduce the symmetric transverse momentum variables,
\begin{equation}
\mathbf{k}_\perp^\pm = \mathbf{k}_\perp \pm \frac{\mathbf{p}_\perp}{2}, 
\end{equation}
and the dimensionless variables
\begin{equation}
x_f = \frac{p_\perp^2}{2|q_f|B}, \qquad \zeta_\pm = \sqrt{\frac{2}{|q_f|B}}\left[\left(k_1 \pm \frac{p_1}{2}\right) + i\left(k_2 \pm \frac{p_2}{2}\right)\right].
\end{equation}

Due to the orthogonality of the Dirac matrices, the cross-terms between the parallel components (\(\gamma^0, \gamma^3\)) and the transverse components (\(\gamma^1, \gamma^2\)) vanish after taking the trace. Therefore, the trace yields
\begin{equation}
\Pi ({\bf p}) = N_c\sum_{f = u,d}\frac{1}{\beta}\sum_n\sum_{l,l'}\int \frac{d^3{\bf k}}{(2\pi)^3}\frac{(-1)^{l+l'}e^{-\frac{|\zeta_+|^2 + |\zeta_-|^2}{2}}{\cal N}_{l,l'}}{\left[(i\omega_n + \mu)^2 - E_{f,l}^2(k_z)\right]\left[(i\omega_n + \mu)^2 - E_{f,l'}^2(k_z+p_z)\right]},
\end{equation}
with
\begin{align}
\nonumber\mathcal{N}_{l,l'} =& 8\left[(i\omega_n + \mu)^2 - k_z(k_z + p_z) + \bar{\sigma}^2\right] \Big[ L_l(|\zeta_-|^2)L_{l'}(|\zeta_+|^2) + L_{l-1}(|\zeta_-|^2)L_{l'-1}(|\zeta_+|^2) \Big] \\
&\quad - 64 \mathbf{k}_\perp^- \cdot \mathbf{k}_\perp^+ L_{l-1}^1(|\zeta_-|^2)L_{l'-1}^1(|\zeta_+|^2).\label{eq:app_trace}
\end{align}

To evaluate the transverse momentum integrals in Eq.~\eqref{eq:app_trace}, we utilize the following identities,
\begin{equation}
\int \frac{d^2{\bf k}_\perp}{(2\pi)^2} e^{-\frac{|\zeta_+|^2 + |\zeta_-|^2}{2}} L_l(|\zeta_+|^2) L_{l'}(|\zeta_-|^2)
= (-1)^{l-l'} \frac{|q_f|B}{8\pi} \frac{l_<!}{l_>!} e^{-x_f} x_f^{|l-l'|} \left[L_{l_<}^{|l-l'|}(x_f)\right]^2,
\end{equation}
and 
\begin{equation}
\int \frac{d^2{\bf k}_\perp}{(2\pi)^2} e^{-\frac{|\zeta_+|^2 + |\zeta_-|^2}{2}}{\bf k}_+\cdot{\bf k}_-L^1_{l-1}(|\zeta_+|^2) L^1_{l'-1}(|\zeta_-|^2)=(-1)^{l+l'}\frac{(|q_f|B)^2}{16\pi}\frac{l_<!}{(l_>-1)!}e^{-x_f}x_f^{|l-l'|}L^{|l-l'|}_{l<}(x_f)L^{|l-l'|}_{l_<-1}(x_f).
\end{equation}

Integrating over \(\mathbf{k}_\perp\), we thus have
\begin{align}
\Pi({\bf p}) = N_c\sum_{f}\frac{ |q_f|B}{2\pi^2} e^{-x_f} \sum_{l,l'=0}^{\infty} x_f^{|l-l'|} \frac{1}{\beta}\sum_n\int dk_z  
  \frac{\left[(i\omega_n + \mu)^2 - k_z(k_z+p_z) + \bar{\sigma}^2\right] J_f(l,l';{\bf p}_\perp) - K_f(l,l';{\bf p}_\perp)}{\left[(i\omega_n + \mu)^2 - E_{f,l}^2(k_z)\right]\left[(i\omega_n + \mu)^2 - E_{f,l'}^2(k_z+p_z)\right]},\label{eq:app_Pi_f}
\end{align}
where the transverse overlap factors \(J_f\) and \(K_f\) are explicitly given by
\begin{equation}
J_f(l,l';{\bf p}_\perp) = \frac{l_<!}{l_>!}\left[L_{l_<}^{|l-l'|}(x_f)\right]^2 + \frac{(l_<-1)!}{(l_>-1)!}\left[L_{l_<-1}^{|l-l'|}(x_f)\right]^2, 
\end{equation}
\begin{equation}
K_f(l,l';{\bf p}_\perp) = 4|q_f|B \frac{l_<!}{(l_>-1)!} L_{l_<}^{|l-l'|}(x_f) L_{l_<-1}^{|l-l'|}(x_f).
\end{equation}

 Using the standard contour integration techniques for the product of two propagators, we can readily carry out the sum over Matsubara frequencies and obtain
\begin{equation}
T\sum_n \frac{1}{{i\omega_n + \mu}^2 - E_{f,l}^2(k_z)} = -A_{f,l}(k_z), 
\end{equation}
\begin{equation}
T\sum_n \frac{1}{\left[(i\omega_n + \mu)^2 - E_{f,l}^2(k_z)\right]\left[{(i\omega_n + \mu)}^2 - E_{f,l'}^2(k_z+p_z)\right]} = \frac{A_{f,l}(k_z) - A_{f,l'}(k_z+p_z)}{2p_z k_z + p_z^2 + 2|q_f|B(l' - l)},
\end{equation}
where \(A_{f,l}(k_z)\) is defined as
\begin{equation}
A_{f,l}(k_z) = \frac{1 - \eta(E_{f,l}(k_z)) - \bar{\eta}(E_{f,l}(k_z))}{2E_{f,l}(k_z)},
\end{equation}
with \(\eta(E)\) and \(\bar{\eta}(E)\) being the Fermi-Dirac distributions for particles and antiparticles. 

Finally, by substituting the Matsubara sums back into Eq.~\eqref{eq:app_Pi_f} and adding the tree-level contribution \(\frac{1}{2G}\), we arrive at the full two-point function,
\begin{align}
\nonumber\Gamma^{(2)}({\bf p}_\perp,p_z) = &\frac{1}{2G} + N_{c}\sum_{f}\frac{|q_{f}|B}{2\pi^{2}} e^{-x_f} \sum_{l,l^{\prime} = 0}^{\infty}x_f^{|l - l^{\prime}|} \int dk_z \Bigg\{-A_{f,l}(k_z)J_f(l,l^{\prime};{\bf p}_\perp) \\
&+\frac{\left(p_z^{2} + p_zk_z + 2|q_{f}|Bl^{\prime} + 2\bar{\sigma}^{2}\right) J_f(l,l^{\prime};{\bf p}_\perp) - K_f(l,l^{\prime};{\bf p}_\perp)}{2p_zk_z + p_z^{2} + 2|q_{f}|B(l^{\prime} - l)} \left[A_{f,l}(k_z) - A_{f,l^{\prime}}(k_z + p_z)\right]\Bigg\} .\label{eq:app_Gamma2_final}
\end{align}
This completes the derivation of eq.~(\ref{eq:B_Gamma2_final}) in Section III.

\bibliography{SC}

\clearpage
%\onecolumngrid

%\newpage 

\end{document}